\documentclass[manuscript]{emulateapj-rtx4} 
\usepackage{natbib}

\usepackage{epsfig,amssymb,amsmath,bm}
\usepackage{color}

\def\bea{\begin{eqnarray}}
\def\eea{\end{eqnarray}}
\def\be{\begin{equation}}
\def\ee{\end{equation}}
\newcommand{\de}{\mathrm d}

\newcommand{\om}{{\Omega_m}}

\newcommand{\ho}{{H_0}}

\begin{document}
\title{A novel approach in the WIMP quest: \\
Cross-Correlation of Gamma-Ray Anisotropies and Cosmic Shear}

\author{Stefano Camera}
%\email{stefano.camera@ist.utl.pt}
\affiliation{CENTRA, Instituto Superior T\'ecnico, Universidade T\'ecnica de Lisboa, Lisboa, Portugal}
\author{Mattia Fornasa}
%\email{fornasam@gmail.com}
\affiliation{School of Physics and Astronomy, University of Nottingham, Nottingham, United Kingdom}
\author{Nicolao Fornengo}
%\email{fornengo@to.infn.it}
\author{Marco Regis}
%\email{regis@to.infn.it}
\affiliation{Dipartimento di Fisica, Universit\`a  di Torino and INFN, Torino, Italy}

%\date{\today}

\begin{abstract}
Both cosmic shear and cosmological gamma-ray emission stem from
the presence of Dark Matter (DM) in the Universe: DM structures are 
responsible for the bending of light in the weak lensing regime and those 
same objects can emit gamma-rays, either because they host astrophysical 
sources (active galactic nuclei or star-forming galaxies) or directly by 
DM annihilations (or decays, depending on the properties of the
DM particle).
Such gamma-rays should therefore exhibit strong correlation with the cosmic
shear signal. In this Letter, we compute the cross-correlation angular power
spectrum of cosmic shear and gamma-rays produced by the annihilation/decay 
of Weakly Interacting Massive Particle (WIMP) DM, as well as by astrophysical 
sources. We show that this observable
provides novel information on the composition of the Extra-galactic
Gamma-ray Background (EGB), since the amplitude and shape of the 
cross-correlation signal strongly depend on which class of sources is 
responsible for the gamma-ray emission. If the DM contribution to the EGB is 
significant (at least in a definite energy range), although compatible with 
current observational bounds, its strong correlation with the cosmic shear 
makes such signal potentially detectable by combining Fermi-LAT data
with forthcoming galaxy surveys, like Dark Energy Survey and Euclid. 
At the same time, the same signal would demonstrate that the weak 
lensing observables are indeed due to particle DM matter and not to possible
modifications of General Relativity.

\end{abstract}


\maketitle

\section{Introduction}
%{\em Introduction.}
Weak gravitational lensing refers to the small distortions of  images of 
distant galaxies, produced by the distribution of matter located between 
galaxies and the observer
\citep{Kaiser:1991qi,Bartelmann:1999yn,Munshi:2006fn,Bartelmann:2010fz}. 
%Contrary to strong lensing, where distortion is caused by {a small number 
%of} massive objects, weak lensing studies aim at reproducing the statistical 
%properties of the density field acting as a lens, as well as investigating the 
%geometrical properties of the Universe. 
A  distorted image can be described 
by the so-called distortion matrix, normally parameterized in terms of  the
convergence $\kappa$ (controlling modifications in the size of the image) and 
the shear $\gamma$ (accounting for shape distortions). {Whilst the former 
is a direct estimator of  matter density fluctuations along the line of 
sight, the latter is easier to measure, through correlations in the observed source 
ellipticities. In the flat-sky approximation, the two generate
identical angular power spectra and we thus focus on the shear.
%as an estimator of the convergence, and, from now on, we indicate with $\kappa$ the 
%observables related to cosmic shear.

The auto-correlation between the gravitational shear {in} two different 
directions can provide information on the clustering of the large scale 
structures responsible for the lensing effect. The technique has already been 
used in \citep{Jullo:2012ty,Tereno:2010dt,delaTorre:2010pt,Schrabback:2009ba} 
where data of the COSMOS galaxy survey allow for a measurement of the 
two-point correlation function of the shear at angular scales between 
0.1 and 10 arcmin. Future surveys, like Pan-STARRS,  Dark Energy Survey (DES) 
\citep{Abbott:2005bi} and  Euclid \citep{Euclid}, due to their larger coverage 
and improved sensitivities, will be able to reconstruct two-dimensional shear 
maps, from which one can extract the auto-correlation angular power spectrum 
(PS).

The same Dark Matter (DM) structures that act as lenses can themselves emit 
light at various wavelengths, including the $\gamma$-ray range. While 
$\gamma$-rays can be produced by astrophysical sources hosted by DM halos 
(i.e. star-forming galaxies (SFG) or active galactic nuclei (AGN)), DM itself may 
be a source of $\gamma$-rays, through its self annihilation or decay, 
depending on the properties of the DM particle. Those $\gamma$-rays emitted 
by DM should therefore have the potential to exhibit strong correlation with 
the gravitational lensing signal.

In this Letter we propose to study the cross-correlation of gravitational 
shear with the Extra-galactic Gamma-ray Background (EGB), i.e. the residual 
radiation contributed by the cumulative emission of {\it unresolved} 
$\gamma$-ray sources, as a novel and potentially relevant channel of DM 
investigation. 

The most recent measurement of the EGB was performed by the Fermi-LAT 
telescope in \citep{Abdo:2010nz}, covering a range between 200 MeV and 100 
GeV: the emission is obtained by subtracting the contribution of resolved  
sources (both point-like and extended) and the Galactic foreground (due to 
cosmic rays interaction with the interstellar medium) from the whole Fermi-LAT 
data. Unresolved astrophysical sources like blazars 
\citep{Abazajian:2010pc,Stecker:2010di,Singal:2011yi}, SFGs
\citep{Fermi:2012eba,Lacki:2012si} or radio galaxies 
\citep{Inoue:2011bm,Massaro:2011ww} contribute to the EGB but {the exact 
amount of their contribution} is still unknown. 
The $\gamma$-rays produced by DM annihilation or decay can also contribute to  
EGB \citep{Ullio:2002pj,Zavala:2009zr,Cirelli:2010xx,Zavala:2011tt,
Fornasa:2012gu}. However, the fact that the EGB energy spectrum is compatible 
with a  power-law, without any evident spectral feature, suggests that 
DM cannot play a leading role in the whole energy range 
\citep{Abdo:2010dk,Calore:2011bt}. 
In the angular anisotropies of the EGB emission, the DM also plays a 
subdominant role: indeed, a detection of a significant auto-correlation 
angular PS has been recently reported \citep{Ackermann:2012uf} (for multipoles 
$\ell>100$, which is the range of interest for our analysis, since there the 
contamination of the Galactic foreground can be neglected), but the features 
of such a signal (in particular its independence on multipole and energy) 
seem to indicate an interpretation in terms of blazars 
\citep{Cuoco:2012yf,Harding:2012gk}.
The contribution of unresolved astrophysical sources to the EGB can also be 
analyzed by cross-correlating the gamma-ray emission with available catalogs 
of resolved galaxies \citep{Xia:2011}.

Both cosmic shear and $\gamma$-ray emission depend on the large scale 
structure of the Universe: because this is what generates the lensing 
effect and because those same structures can produce $\gamma$-rays, either from the hosted
astrophysical sources or directly from DM annihilation/decay. A certain level 
of cross-correlation between cosmic shear and $\gamma$-ray emission is 
therefore expected. The key point of our analysis is to 
understand whether the shear/$\gamma$-rays cross-correlation is within reach 
of future galaxy surveys, as DES and Euclid, and under which circumstances 
such signal can be proficiently used to disentangle a true DM signal from the 
other astrophysical $\gamma$-rays sources. It is the first time, to our 
knowledge, that such analysis is performed. If successful, this approach could 
provide direct evidence that what is measured by weak lensing surveys is indeed 
due to DM and is not, e.g., a manifestation of alternative theories of 
gravity. 

\begin{figure}[t!]
 \begin{minipage}[htb]{0.225\textwidth}
   \centering
   \includegraphics[width=\textwidth]{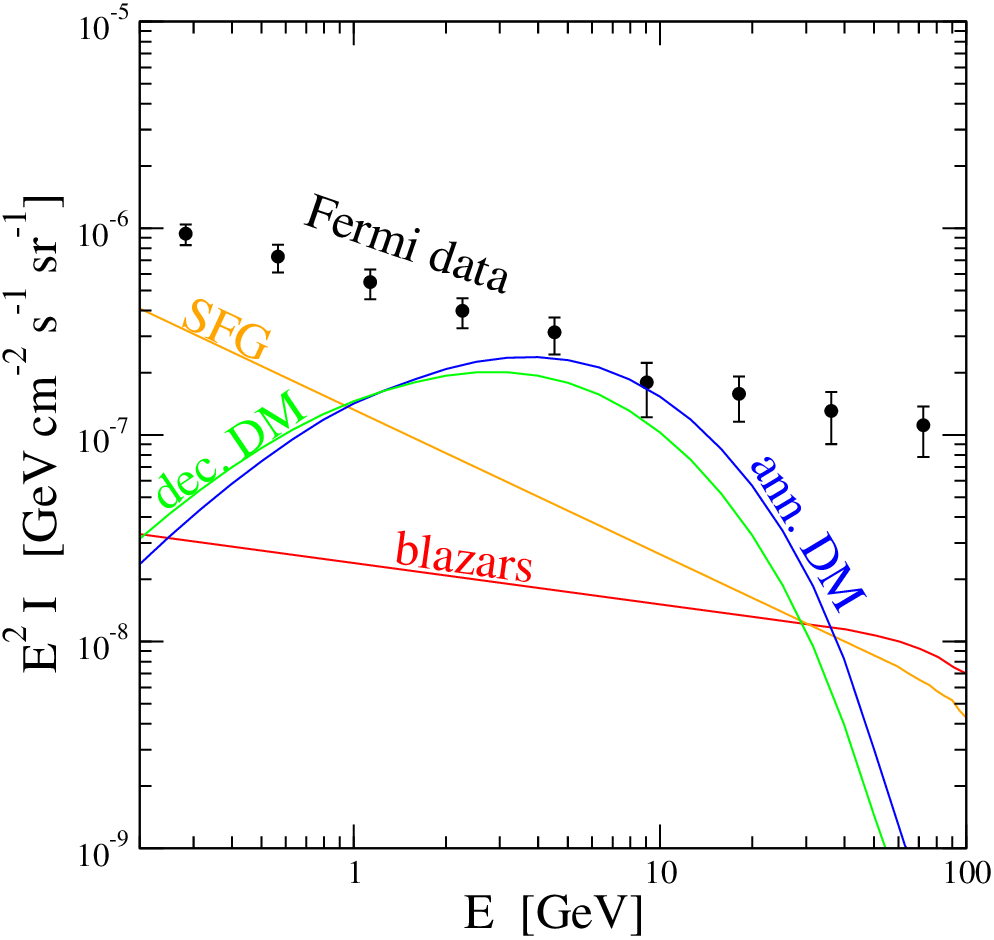}
 \end{minipage}
 \ \hspace{0.5mm} \
 \begin{minipage}[htb]{0.225\textwidth}
   \centering
   \includegraphics[width=\textwidth]{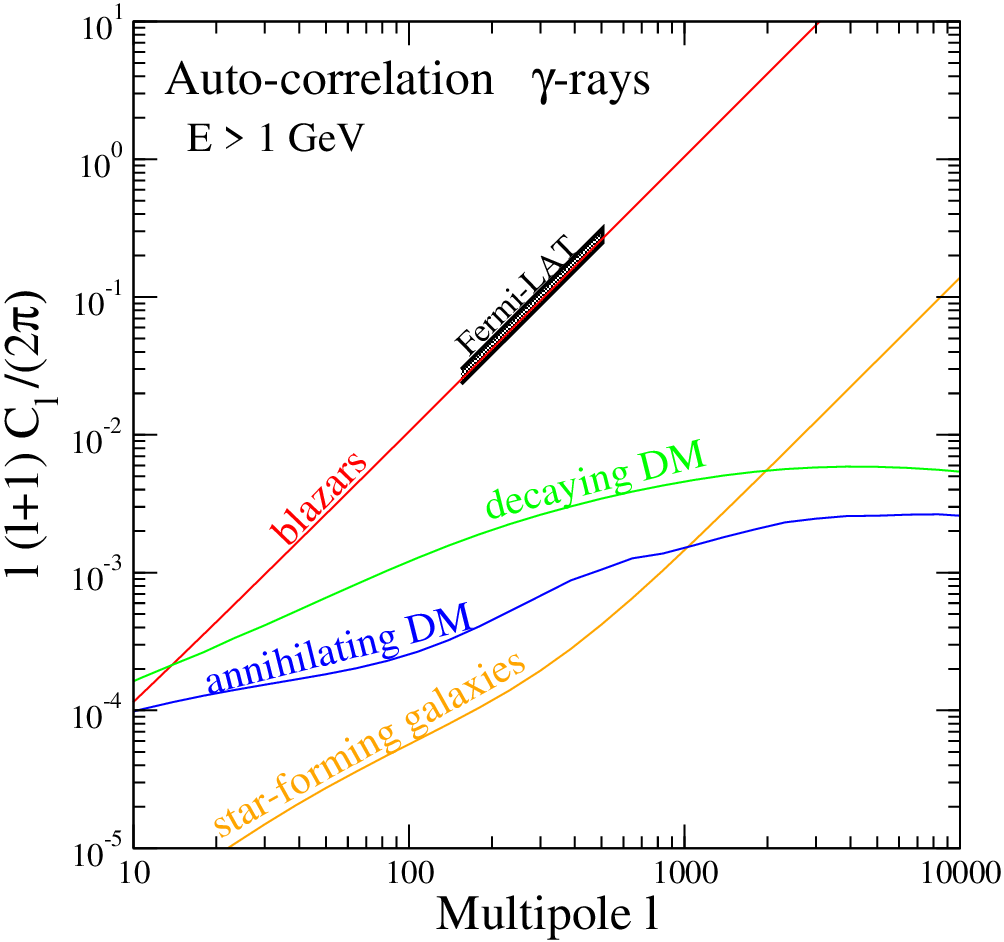}
 \end{minipage}
    \caption{{\it Left}: EGB emission as a function of observed energy for the four extragalactic components described in the text. Data are from \citet{Abdo:2010nz}. {\it Right:} $\gamma$-ray angular PS at $E>1$ GeV for the same models of the left panel. {The observed angular PS is summarized by the black band \citep{Ackermann:2012uf}}.}
\label{fig:gamma}
 \end{figure} 

 \begin{figure}[ht!]
 \begin{minipage}[htb]{0.225\textwidth}
   \centering
   \includegraphics[width=\textwidth]{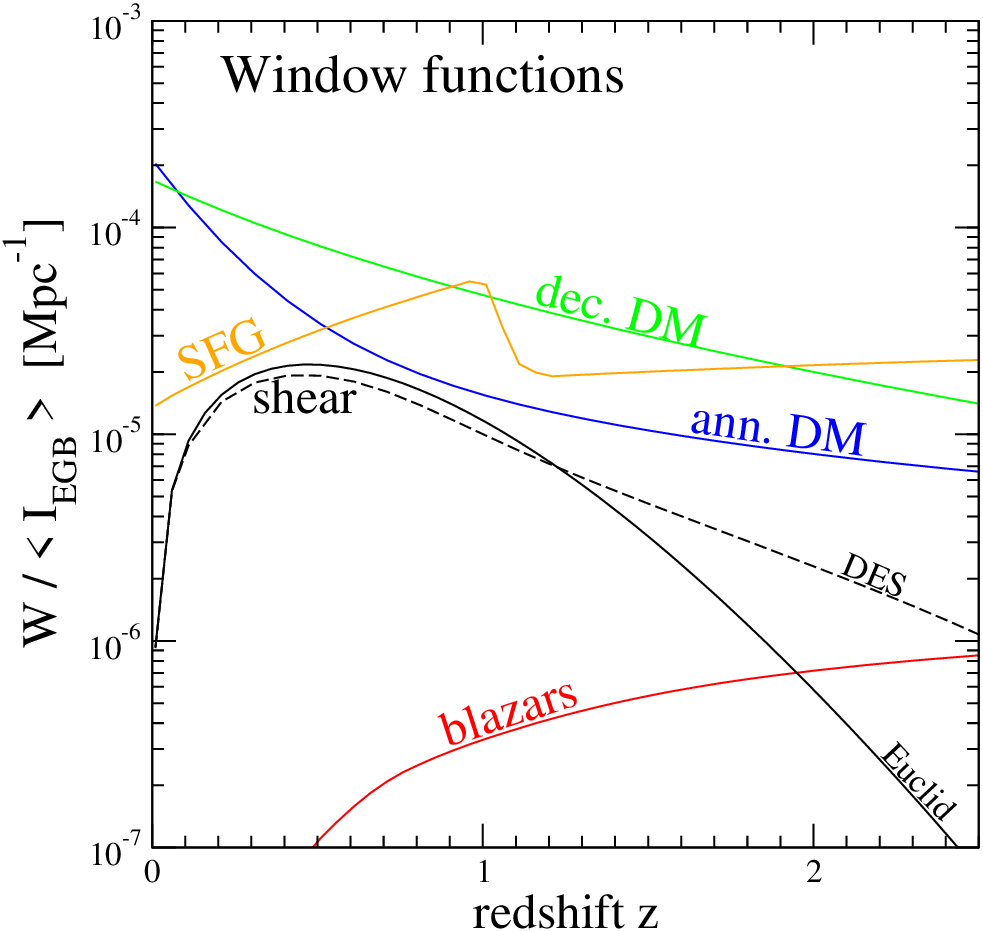}
 \end{minipage}
 \ \hspace{0.5mm} \
 \begin{minipage}[htb]{0.225\textwidth}
   \centering
   \includegraphics[width=\textwidth]{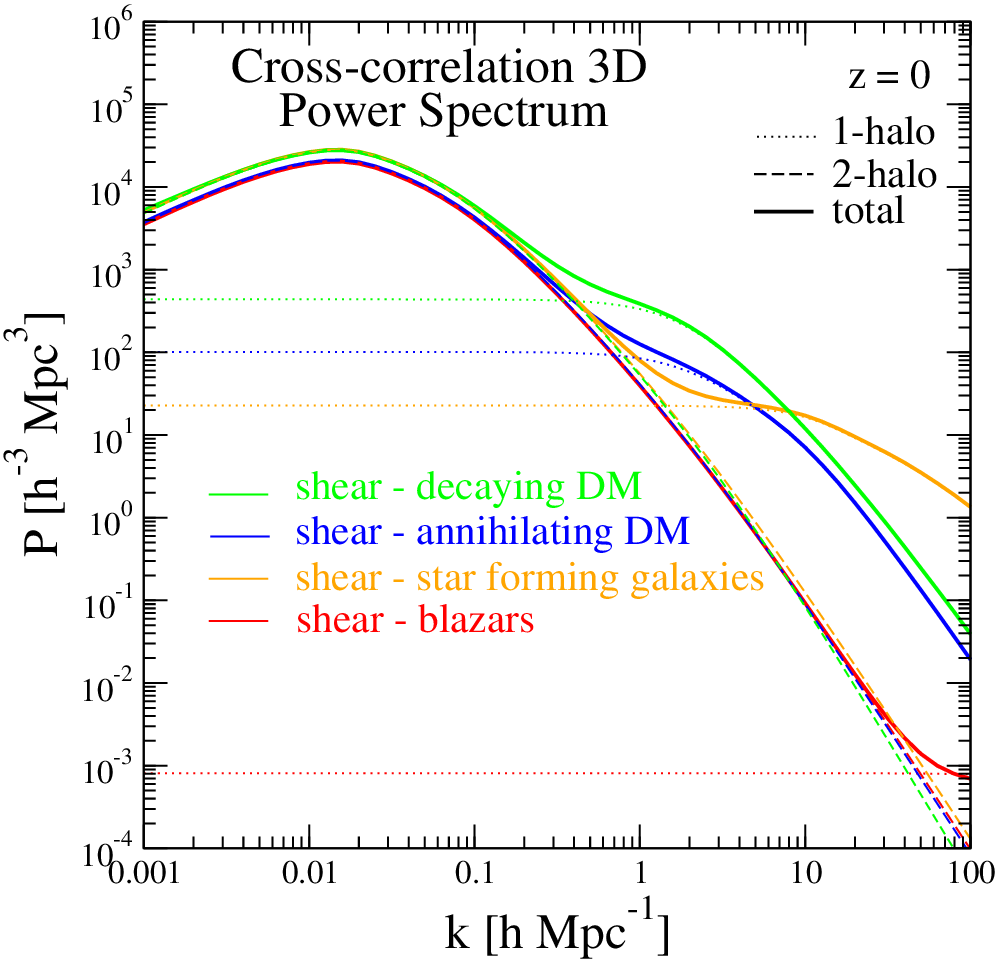}
 \end{minipage}
\caption{{\it Left}: Window functions vs. redshift. For $\gamma$-ray sources we consider the flux above 1 GeV normalized to the total EGB intensity measured by Fermi-LAT. {\it Right:} Three-dimensional PS of  cross-correlation shear/$\gamma$-rays at $z=0$. }
\label{fig:crossmod}
 \end{figure}

\section{Theoretical modeling} 
%{\em Theoretical modeling.} 
%
\begin{figure*}[t]
 \centering
 \includegraphics[width=0.30\textwidth]{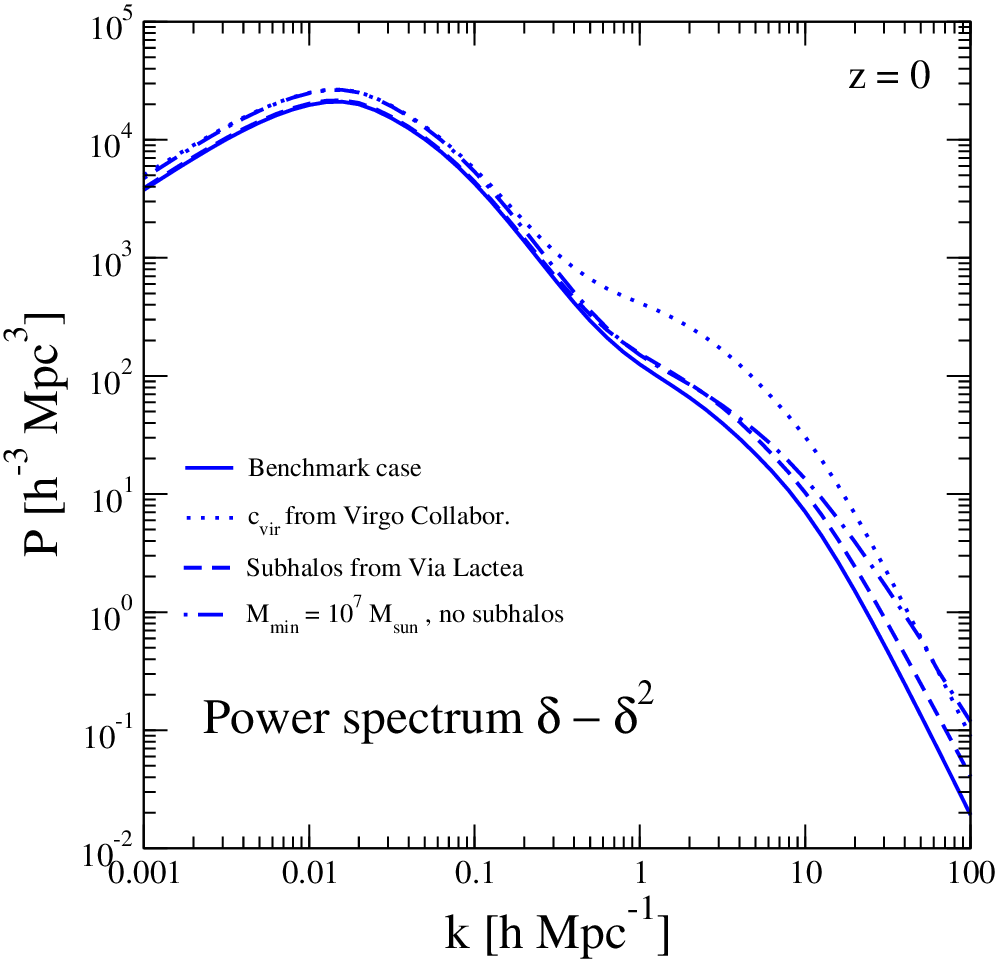}~
 \includegraphics[width=0.30\textwidth]{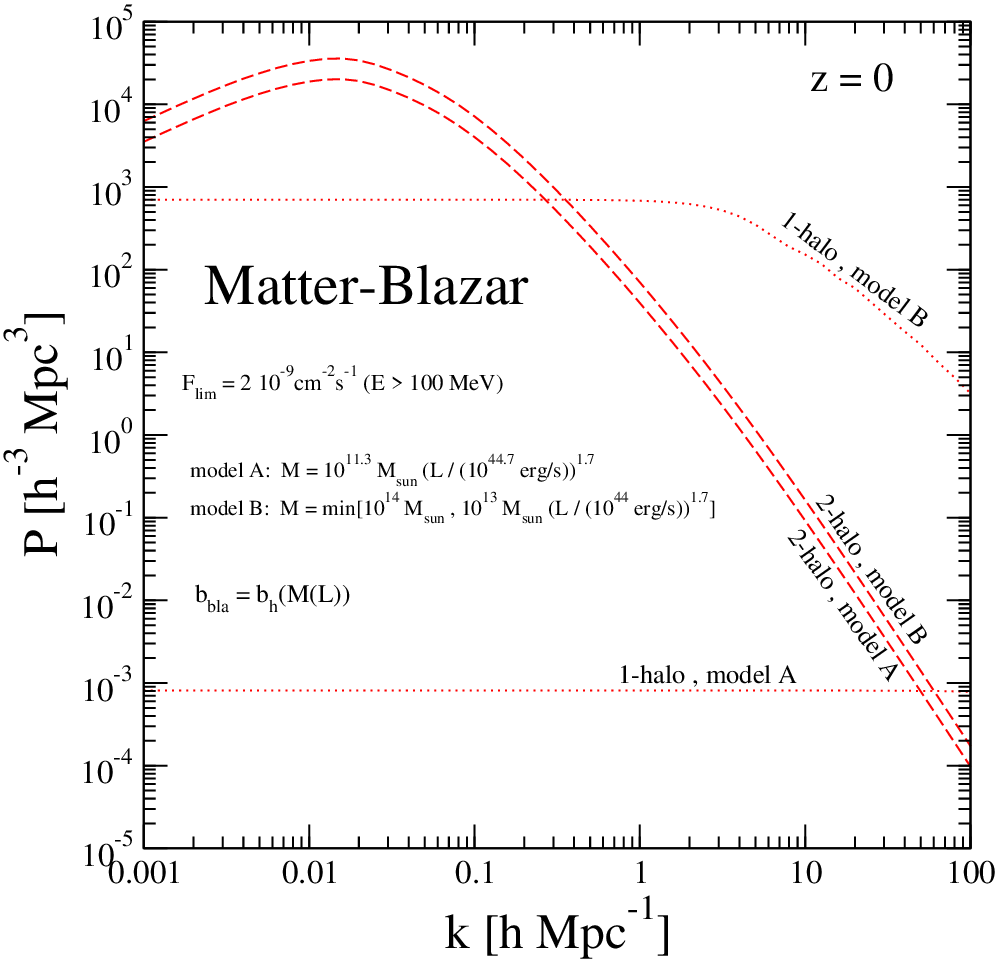}~
 \includegraphics[width=0.30\textwidth]{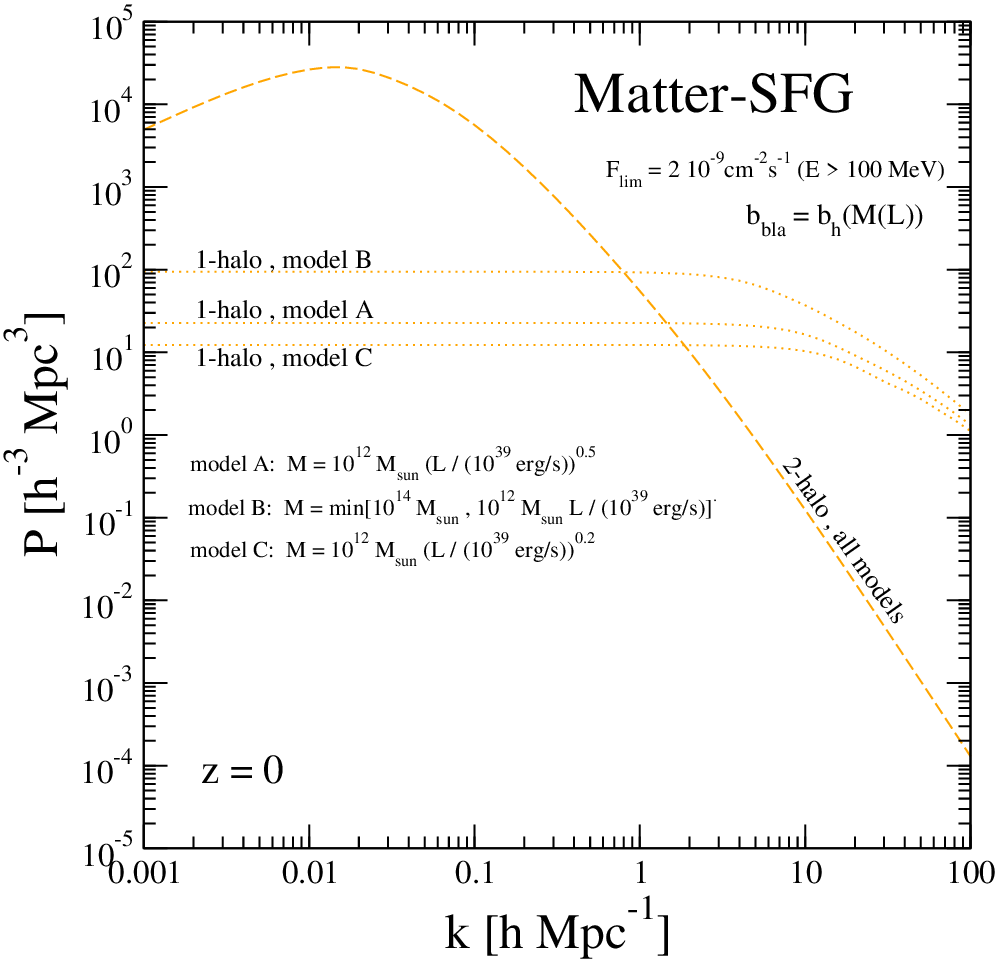}
 \caption{{\it Left}: 3D cross-correlation PS between shear and annihilating DM in the scenarios described in the text at $z=0$. 
{\it Central:} 3D cross-correlation PS between shear and $\gamma$-rays from blazars at $z=0$, with two models for 
$m(\mathcal{L})$ \citet{Ando:2006mt} ($\mathcal{L}$ is the $\gamma$-ray luminosity at 100 MeV).  
{\it Right:} Same as central panel but for SFGs ($\mathcal{L}$ now being the $\gamma$-ray luminosity between 100 MeV and 100 GeV). 
}
\label{fig:pscross}
 \end{figure*}

The source intensity along a given direction 
$\vec n$ can be written as:
\be
 I_g (\vec n) = \int d\chi\, g(\chi,\vec n)\,\tilde W(\chi)\;,
  \label{eq:int}
\ee
where $\chi(z)$ is the radial comoving distance, $g$ is the density field of 
the source, and $\tilde W$ is the window function (which does not depend on 
$\vec n$). We then define a normalized version 
$W=\langle g \rangle\,\tilde W$, so $\langle I_g \rangle = \int d\chi\ W(\chi)$.
Expanding the intensity fluctuations of two source populations $i$ and $j$ in 
spherical harmonics, one can compute the cross-correlation 
angular PS (here in the dimensionless form):
\be
 C_\ell^{(ij)} = 
 \frac{1}{\langle I_i \rangle\langle I_j \rangle} \int \frac{d\chi}{\chi^2} W_i(\chi)\, W_j(\chi)
 P_{ij}(k=\ell/\chi,\chi)\;.
  \label{eq:Cl}
\ee
The definition of the 3-dimensional PS $P_{ij}$ is 
$\langle \hat f_{g_i} (\chi,\bm k) \hat f_{g_j}^\ast (\chi^\prime,\bm k^\prime) \rangle = (2\pi)^3 \delta^3 (\bm k - \bm k^\prime) P_{ij}(k,\chi,\chi^\prime)$, 
where $f_g \equiv [g(\bm x|m,z)/\bar g(z)-1]$ ($\hat f_g$  is its Fourier transform) and the Limber 
approximation ($k=\ell/\chi$) is assumed to hold. We consider the sources to 
be characterized by a parameter $m$ (typically the mass), and $g(\bm x|m)$ is 
the density field of an object associated to $m$, while 
$\bar g(z)=\langle g(\vec n,z)\rangle$.
$P_{ij}$ can be computed following the so-called halo-model approach. The 
two-point correlation is given by the sum of two components, the 1-halo and 
2-halo terms, i.e. $P_{ij}=P_{ij}^{1h}+P_{ij}^{2h}$ 
\citep{Scherrer:1991,Ando:2005xg,Ando:2006cr}:

\bea
 P_{ij}^{1h}(k) &=& \int dm\ \frac{dn}{dm} \hat f_i^\ast(k|m)\,\hat f_j(k|m) \label{eq:PS1halo}\\
 P_{ij}^{2h}(k) &=& \left[\int dm_1\,\frac{dn}{dm_1} b_i(m_1) \hat f_i^\ast(k|m_1) \right] \nonumber\\ 
                &\times& \left[\int dm_2\,\frac{dn}{dm_2} b_j(m_2) \hat f_j(k|m_2) \right]\,P^{\rm lin}(k)\;,\label{eq:PS2halo}
\eea
where $dn/dm$ is the number density distribution of sources, $P^{\rm lin}$ is 
the linear matter PS, and $b_i(m)$ is the linear bias between the object $i$ 
and matter.

Note that the average of $\langle g \rangle$ is given by:
\be
\bar g(z)=\langle g(\vec n,z) \rangle=\int \ dm\,\frac{dn}{dm}\int\ d^3\bm x\,g(\bm x|m,z)\;,
\label{eq:gave}
\ee
which implies that at small $k$ (where 
$\hat f\sim\int d^3\bm x\,g(\bm x|m)/\bar g $) the terms in the 
square-brackets in Eq.~(\ref{eq:PS2halo}) are $\sim1$ (except in the case of 
a significant bias). The 2-halo term is thus normalized to the standard linear 
matter PS at small $k$, which motivates the normalization of the window 
function introduced above.

We aim at cross-correlating the shear signal (source $i$ in 
Eqs. (\ref{eq:Cl}--\ref{eq:PS2halo})) with $\gamma$-rays emitted by DM, 
SFGs, and blazars (source $j$ in Eqs.  (\ref{eq:Cl}--\ref{eq:PS2halo})). 
For what concerns weak lensing, $W$ takes the form (see, e.g.,~\citep{Bartelmann:2010fz}):
\be
W^\kappa(\chi)=\frac{3}{2}\ho^2\om[1+z(\chi)]\chi\int_\chi^\infty\!\!\de\chi'\,\frac{\chi'-\chi}{\chi'}\frac{\de N}{\de\chi'}(\chi')
\nonumber\label{eq:W(z)}
\ee
and $\de N/\de \chi$ represents the redshift distribution of the sources, 
normalized to unity area (such that we can take $\langle I_i \rangle=1$ in 
Eq.~(\ref{eq:Cl})).
For Euclid  $\de N/\de z=A_E\,z^2\,e^{-(z/z_0)^{1.5}}$, where $z_0=z_m/1.4$ with 
$z_m=0.9$ being the median redshift of the survey and $A_E$ is fixed by the 
normalization $\int dz \de N/\de z=1$. 
For DES, $\de N/\de z=A_D\,(z^a+z^{ab})/(z^b+c)$, with $a$, $b$, and $c$ 
provided in Table 1 of \citep{Fu:2007qq}, and $A_D$ fixed by the normalization.
Since gravitational lensing is sourced by the potential wells of the large 
scale structure, whose Poisson equation relates to the matter distribution 
$\rho$, we have $g_\kappa(\bm x)=\rho(\bm x)$, and $f_\kappa(\bm x)$ is given by 
the density contrast $\delta(\bm x)$. For the bias in Eq. (\ref{eq:PS2halo})  
we use the estimates in \citep{Cooray:2002dia}.
We adopt the halo mass function $dn/dm$ of \citep{Sheth:1999mn}, the halo
concentration from \citep{MunozCuartas:2010ig}, and a NFW halo density 
profile~\citep{Navarro:1996gj}.

For the case of $\gamma$-rays from decaying DM we again have 
$f_d(\bm x)=\delta(\bm x)$ (we assume $\rho_m\simeq\rho_{DM}$). The window 
function is now given by:
\be
W^{\gamma_d}(E_\gamma,z) = \frac{1}{4\pi} \frac{\Omega_{DM}\rho_c}{m_\chi \tau_d} J_d(E_\gamma,z)\;,
\ee
where $m_\chi$ and $\tau_d$ are the mass and decay lifetime
of the DM particle, $J_d=\int_{E_\gamma}^\infty dE \frac{dN_d(E(1+z))}{dE} e^{-\tau(E(1+z),z)}$ 
with $dN_d/dE(E)$ being the number of $\gamma$-ray photons emitted per decay event
in $(E,E+dE)$, and $\tau$ being the optical depth for 
absorption~\citep{Stecker:2006eh}.
% mainly due to pair production on the extragalactic background light emitted by galaxies in the ultraviolet, optical, and infrared bands.
Note that the factor $\Omega_{DM}\rho_c$ comes from the normalization of $W$, 
since in this case $\langle g_d \rangle=\bar \rho$.

The DM annihilation signal scales with $\rho^2$, thus we have 
$\hat f_{a}\equiv \tilde u(k|m)$ given by the Fourier transform of 
$\rho^2(\bm x|m)/\langle \rho^2 \rangle$.
In the literature, equations are often written in terms of the so-called 
clumping factor:
\be
\Delta^2(z)=\frac{\langle \rho^2 \rangle}{\bar \rho^2}=\int_{m_{\rm min}}^{m_{\rm max}} dm\,\frac{dn}{dm}\int\ d^3\bm x\,\frac{\rho^2(\bm x|m)}{\bar \rho^2}\;,
\label{eq:clumpfact}
\ee
and the window function has the form:
\be
W^{\gamma_a}(E_\gamma,z) = \frac{(\Omega_{DM}\rho_c)^2}{4\pi} \frac{(\sigma_a v)}{2 m_\chi^2} (1+z)^3\,\Delta^2(z)\,J_a(E_\gamma,z)
\;,
\nonumber\label{eq:wann}
\ee
where $(\sigma_a v)$ is the velocity-averaged annihilation rate (which we 
 assume to be the same in all halos) and 
$J_a=\int_{E_\gamma}^\infty dE \frac{dN_a}{dE}(E(1+z)) \,e^{-\tau(E(1+z),z)}$ with 
$dN_a/dE(E)$ being the number of $\gamma$-ray photons emitted per annihilation 
in the energy range $(E,E+dE)$.
In the annihilating DM case, the predictions for both the window function and 
the PS heavily depend on the (unknown) clustering at small masses (i.e., on 
the minimum halo mass, concentration below approximately $10^6M_\odot$, and on
the amount of substructures). In our benchmark case, we consider $m_{\rm min}=10^{-6}M_\odot$ (typical 
free-streaming mass for WIMPs) and include unresolved subhalos following the 
scheme of \citep{Kamionkowski:2010mi} with parameters tuned as in 
the HIGH scenario of Sec. 3.3 in \citep{Fornasa:2012gu} (within our halo model, 
it  induces only moderate boost factor $\sim2$).
In Fig.~\ref{fig:pscross}c and \ref{fig:apscross}c, 
we estimate the impact of clustering uncertainties by considering a different subhalo 
scheme (from Via Lactea \citep{Kamionkowski:2010mi}), 
a different extrapolation of the halo concentration (from Virgo Collaboration \citep{Fornasa:2012gu}), 
and a conservative case with $m_{\rm min}=10^{7}M_{\odot}$ 
(minimum halo mass currently inferred from dynamical measurements) 
and no substructures.

\begin{figure*}[t]
 \centering
 \includegraphics[width=0.30\textwidth]{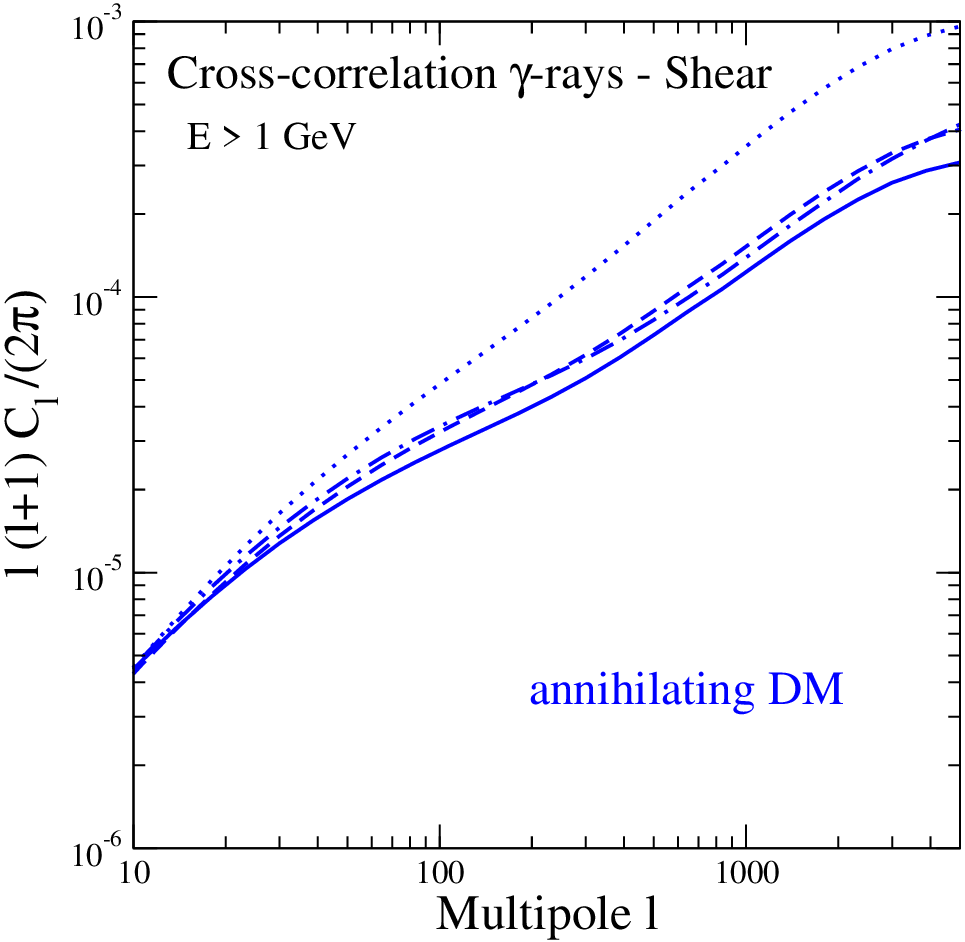}~\;
 \includegraphics[width=0.30\textwidth]{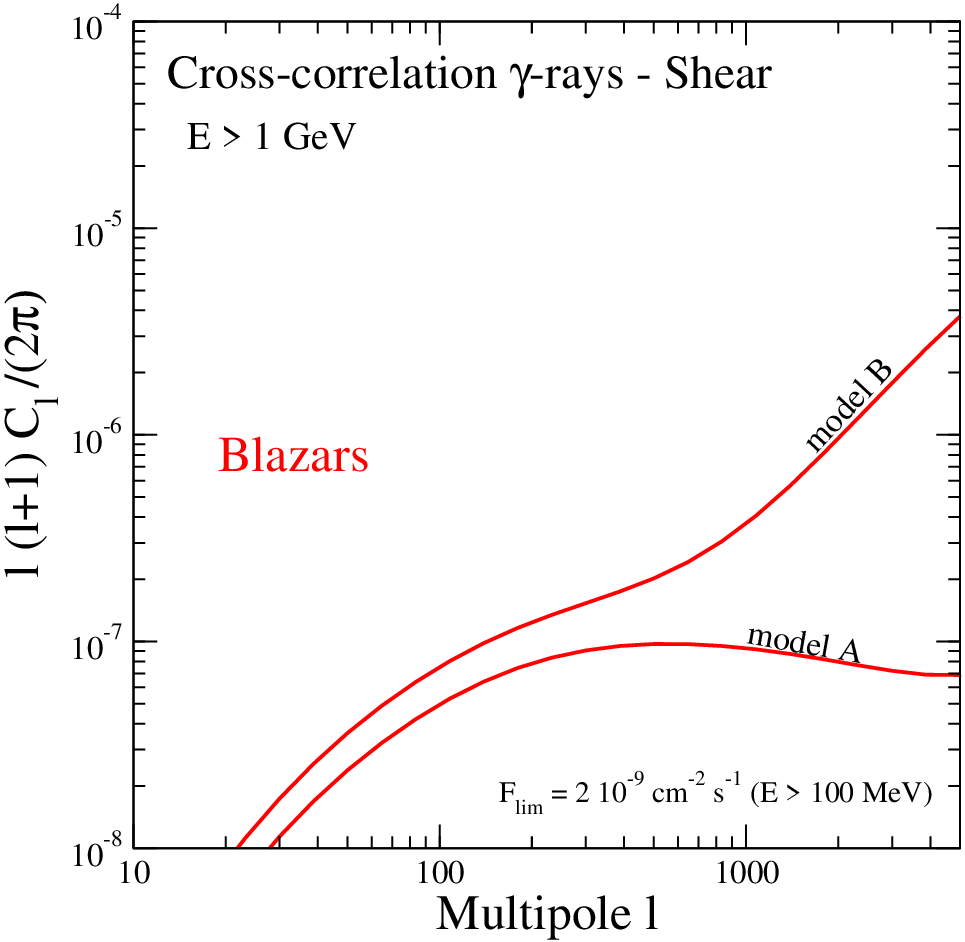}~\;
 \includegraphics[width=0.30\textwidth]{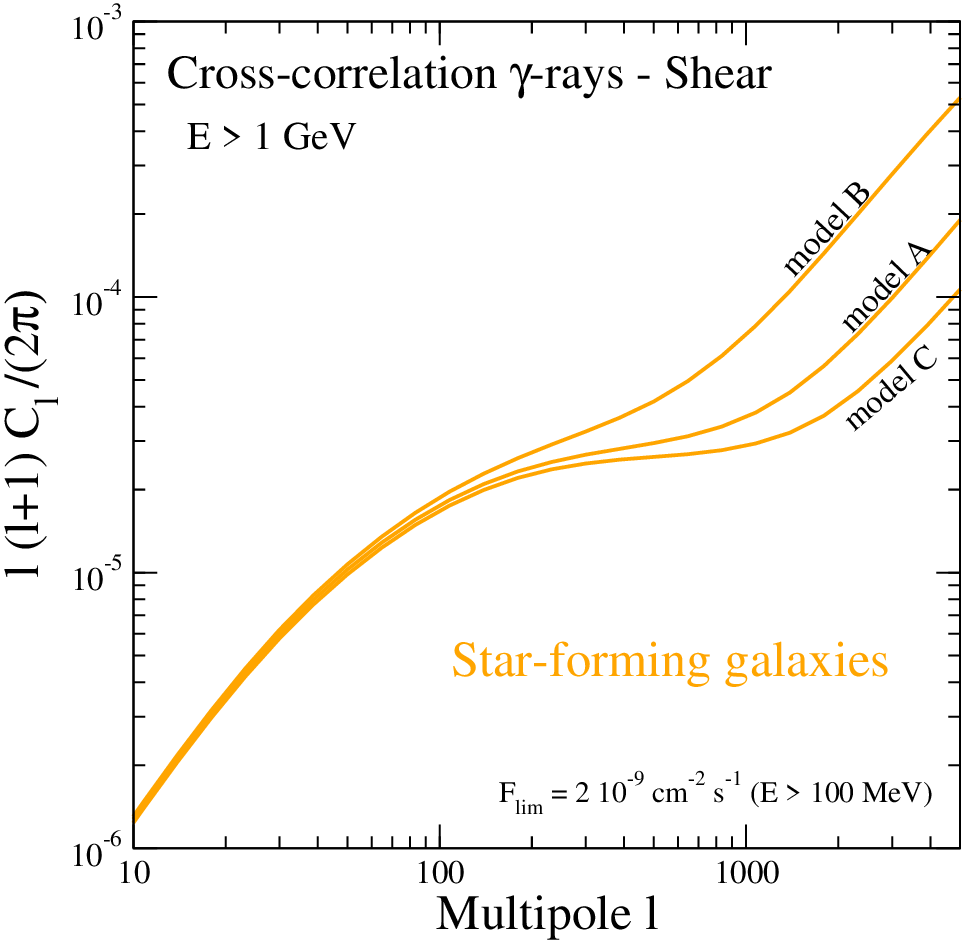}
 \caption{Angular cross-correlation PS between shear and $\gamma$-rays, taking the window functions described in the text and the 3D PS models reported in Fig.~\ref{fig:pscross}. To be compared with Fig.~\ref{fig:crossexp}b.}
\label{fig:apscross}
 \end{figure*}

The formalism sketched in Eqs. (\ref{eq:int}--\ref{eq:gave}) can be used also 
for a population of astrophysical sources, by replacing the mass with the 
source luminosity $\mathcal{L}$ as the characterizing parameter. This leads 
to the replacement of $dm \,dn/dm$ with $d\mathcal{L}\, \Phi$, where $\Phi$ 
is the $\gamma$-ray luminosity function (GLF). For the range of multipoles of 
interest ($\ell<10^3$) both blazars and SFGs can be approximated as point 
sources and we have 
$g_S(\mathcal{L},\bm x-\bm x')=\mathcal{L}\,\delta^3(\bm x-\bm x')$, which 
leads to:
\bea
 P_{\kappa\gamma_S}^{1h}(k,z) &=& \int_{\mathcal{L}_{\rm min}(z)}^{\mathcal{L}_{\rm max}(z)} d\mathcal{L}\,\Phi(\mathcal{L},z)\,\frac{\mathcal{L}}{\langle g_S \rangle} \,\tilde v(k|m(\mathcal{L}))  \nonumber\\
 P_{\kappa\gamma_S}^{2h}(k,z) &=& \left[\int_{\mathcal{L}_{\rm min}(z)}^{\mathcal{L}_{\rm max}(z)} d\mathcal{L}\,\Phi(\mathcal{L},z)\, b_S(\mathcal{L},z)\,\frac{\mathcal{L}}{\langle g_S \rangle} \right]\nonumber \\
&\times&\left[\int_{m_{\rm min}}^{m_{\rm max}} dm\,\frac{dn}{dm} \tilde v(k|m) \right] \,P^{\rm lin}(k,z)\;,\label{eq:PSBd}
\eea
with $\tilde v(k|m)$ being the Fourier transform of $\rho(\bm x|m)/\bar \rho$ 
and $\langle g_S \rangle= \int d\mathcal{L}\,\Phi\,\mathcal{L}$. In 
Eqs. (\ref{eq:PSBd}) a relation between the source luminosity $\mathcal{L}$ and 
the host-halo mass $m$ is required.
We compute the source bias $b_S$ through the halo bias by means of 
$b_S(\mathcal{L},z)=b_h(m_h(\mathcal{L}),z)$, for which we need again a 
relation between host-halo mass and source luminosity. On the other hand, 
since at low redshift and in the mass-range of interest $b_h\sim1$, the 2-halo 
term is only very mildly dependent on the description of $m_h(\mathcal{L})$.
For a power-law spectrum with index $\alpha$, the window function is:
\be
W^{\gamma_S}(E_\gamma,z) = \frac{A_S(z)\,\langle g_S(z) \rangle}{4\pi\,E_0^2}\int_{E_\gamma}^\infty dE \left(\frac{E}{E_0}\right)^{-\alpha} e^{-\tau(E,z)}\;,
\nonumber
\ee
where $E_0=100$ MeV and $A_S$ is a factor that depends on which specific 
luminosity is chosen as the characterizing parameter (as we will describe 
below).

The GLF of blazars is computed following the model described in 
\citep{Inoue:2008pk} with the AGN X-ray luminosity function from 
\citep{Ueda:2003yx} and with the numerical value of parameters derived in 
\citep{Harding:2012gk} by fitting Fermi-LAT data on EGB diffuse emission and 
anisotropies.
%\footnote{With this approach, we can in principle include different sub--populations, like BL Lacertae and Flat Spectrum Radio Quasars, in a single effective description.}
The spectrum is taken to be a power-law with $\alpha=2.2$, and $\mathcal{L}$ 
is the $\gamma$-ray luminosity at 100 MeV (which leads to $A_S=(1+z)^{-\alpha}$).
We assume that no blazars fainter than the luminosity cutoff 
$\mathcal{L}_{\rm min}=10^{42}$ erg/s can exist at any redshift, while 
$\mathcal{L}_{\rm max}(z)$ is the maximum luminosity above which a blazar can 
be resolved (for 5-yr Fermi-LAT it is computed taking 
$F_{\rm max}=2\cdot10^{-9}\,{\rm cm^{-2}s^{-1}}$ for $E>100$ MeV).
The relation between halo-mass and blazar luminosity can be described through 
$m_h=10^{11.3}M_\odot(\mathcal{L}/10^{44.7}{\rm erg/s})^{1.7}$ following 
\citep{Ando:2006mt} where the blazar $\gamma$-ray luminosity is linked to 
the mass of the associated supermassive black hole, which is in turn related 
to the halo mass.
The description of $m_h(\mathcal{L})$ suffers from sizable uncertainties 
which propagate to the prediction of the 1-halo term. However, as 
can be seen from Figs.~\ref{fig:pscross}a and \ref{fig:apscross}a, 
where we introduce an alterantive model (model B) 
which dramatically increases $m_h(\mathcal{L})$ 
with respect to our benchmark case (model A), 
the blazar contribution remains largely subdominant.

For the GLF of SFGs, we follow results from the Fermi-LAT 
Collaboration~\citep{Fermi:2012eba}, which are based on the infrared (IR) 
luminosity function derived in \citep{Rodighiero:2009up}, and the rescaling 
relation between $\gamma$-ray and IR luminosity obtained analyzing resolved 
SFGs~\citep{Fermi:2012eba}.
The spectrum is assumed to be a power-law with $\alpha=2.7$, similar to the 
Milky Way case, and $\mathcal{L}$ is the $\gamma$-ray luminosity between 
0.1 and 100 GeV (which leads to $A_S=(\alpha-2)/(1+z)^2$). 
The dependence of the SFG--shear PS on the $m(\mathcal{L})$ relation is milder than for blazars.
In this case, the relation could, in principle, be computed from the relation between 
$\gamma$-ray luminosity and star formation rate (SFR) \citep{Fermi:2012eba}, 
the Schmidt-Kennicutt law (connecting SFR and gas density), and the ratio of 
gas to total galactic mass. This leads to different relations for each 
different subpopulation of SFGs (e.g., ellipticals are much brighter than 
spirals of the same mass); on the other hand we do not have $\gamma$-ray data 
to compute the specific GLF of the sub-populations, thus we have to derive 
an effective averaged relation. Assuming a power-law scaling 
$m={\cal A}\cdot10^{12}M_\odot(\mathcal{L}/10^{39}{\rm erg/s})^{\cal B}$ and a 
maximum galactic mass of $m_{\rm max}=10^{14}M_\odot$, we can find $\cal A$ and 
$\cal B$ using, e.g., the Milky Way data ($m\simeq10^{12}M_\odot$ and 
$\mathcal{L}\simeq 10^{39}{\rm erg/s}$) and requiring that the mass associated 
to the maximum luminosity $\sim 10^{43}{\rm erg/s}$ (this can be computed from 
the maximum observed IR luminosity~\citep{Rodighiero:2009up} rescaled to $\gamma$-ray frequency~\citep{Fermi:2012eba}) not
to exceed $m_{\rm max}$.
We found ${\cal A} \simeq 1$ and ${\cal B} \simeq 0.5$. This is just a simple
 benchmark model, and we estimated the impact of the associated uncertainty 
(by varying $\cal A$ and $\cal B$ within reasonable ranges) in 
Figs.~\ref{fig:pscross}b and \ref{fig:apscross}b.

\section{Results}
%{\em Results.}
%
For the sake of clearness we focus on a benchmark annihilating (decaying) DM 
scenario, where the WIMP has a mass of 100 GeV (200 GeV), annihilation 
(decay) rate of $(\sigma_a v)=8\cdot 10^{-26}{\rm cm^3/s}$ 
($\tau_d=3\cdot 10^{26}$ s) and dominant final state $\bar{b}b$. {The
characteristics of the DM particle are chosen to saturate (at least in one
particular energy range) the EGB emission, without violating the experimental
constraints.}\footnote{The annihilation rate is degenerate with the clumping factor in setting the size of the signal: different clustering schemes providing larger boost-factors could accommodate smaller values of $(\sigma_a v)$, still obtaining similar predictions for the angular PS.}
In particular, we note that, although we take DM to be a significant 
component of the EGB at $E\gtrsim1$ GeV in Fig.~\ref{fig:gamma}a, it is 
basically impossible to obtain an evidence for DM from the angular PS of 
$\gamma$-rays alone because the latter is dominated by the blazar contribution.

 \begin{figure}[t]
 \begin{minipage}[htb]{0.225\textwidth}
   \centering
   \includegraphics[width=\textwidth]{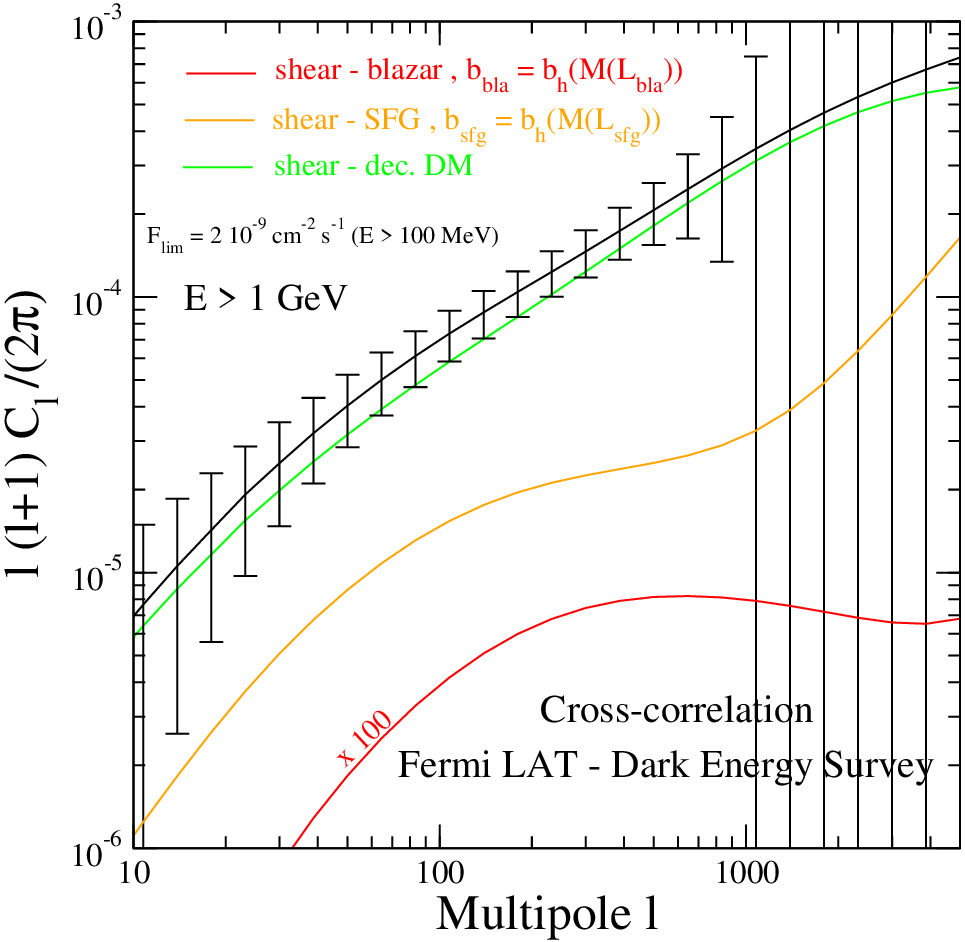}
 \end{minipage}
 \ \hspace{0.5mm} \
 \begin{minipage}[htb]{0.225\textwidth}
   \centering
   \includegraphics[width=\textwidth]{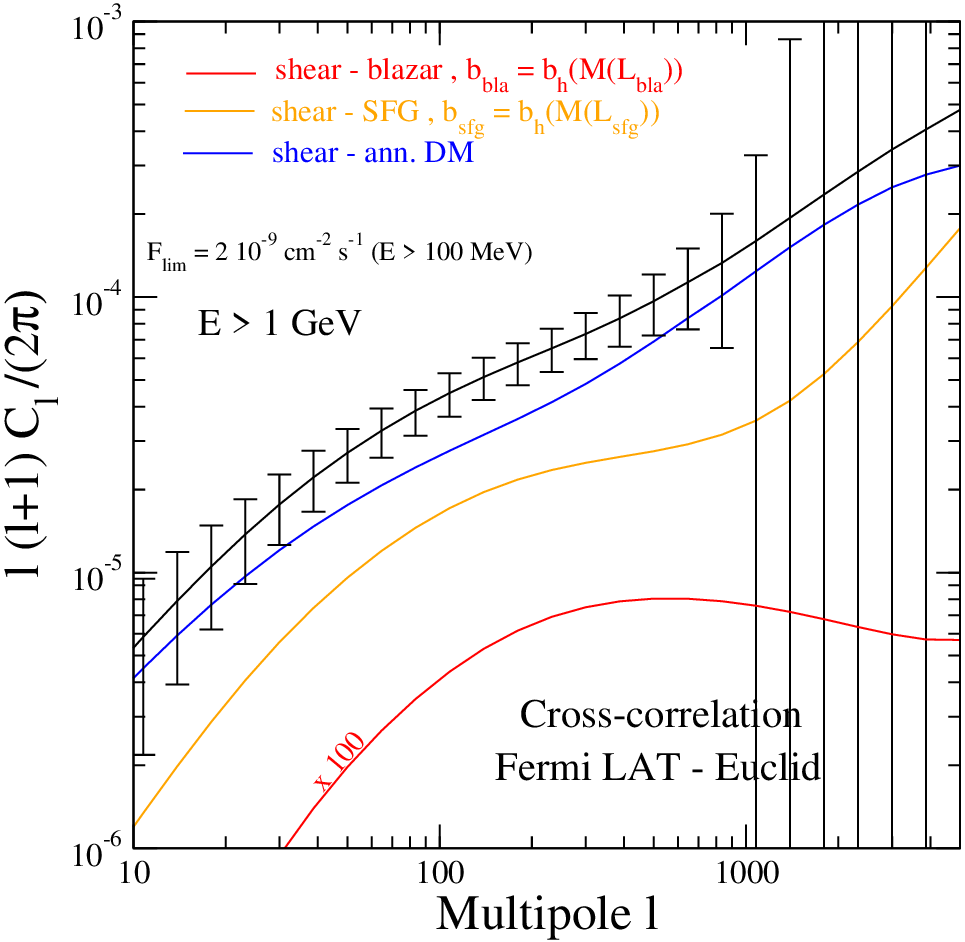}
 \end{minipage}
    \caption{{\it Left}: {Cross-correlation between cosmic shear and $\gamma$-ray emission, for the different classes of $\gamma$-ray emitters described in the text (with a $\gamma$-ray threshold expected for Fermi-LAT after 5 years of exposure). Each contribution is normalized by multiplying Eq. (\ref{eq:Cl}) by $\langle I_j \rangle/\langle I_{\rm EGB} \rangle$ to make them additive. DES is taken as the reference galaxy survey. Error bars are estimated for the total signal (in black). {\it Right:} Same as in the left panel but for annihilating DM, with Euclid as the reference galaxy survey.}}
\label{fig:crossexp}
 \end{figure} 

In Fig.~\ref{fig:crossmod} we show the ingredients of Eq. (\ref{eq:Cl}) for 
the computation of the shear/$\gamma$-ray cross-correlation angular PS: 
{the window function for the cosmic shear signal nicely overlaps  with the
DM window function, both for annihilating and decaying DM, while this happens
only at intermediate redshifts for the SFG window function and only at high redshifts
 for the case of blazars. This suggests 
that a tomographic approach could be a powerful strategy to further disentangle 
different contributions in the angular PS} (this will be pursued in a future 
work~\citep{wip}).
The shear signal is stronger for larger DM masses. The same is true also for the 
$\gamma$-ray signal from DM and this fact gives a large 1-halo contribution 
which dominates starting from $k\lesssim1\, h/{\rm Mpc}$ in 
Fig.~\ref{fig:crossmod}b. 
Galaxies have masses $\lesssim10^{14}M_\odot$, thus they correlate with the 
shear signal of lower-mass halos and the 1-halo contribution becomes important 
at slightly smaller scale $k \gtrsim 1 h/{\rm Mpc}$.
Since the bulk of unresolved blazars in 5-yr Fermi-LAT will be hosted in 
relatively small halos at large redshift, the 1-halo term of the blazar/shear 
PS is suppressed.
Thus, an important result is that, since both the shear and DM-induced 
$\gamma$-ray signals are stronger for larger halos, their cross-correlation is 
more effective with respect to the case of astrophysical sources.
This, together with the sizable overlapping of the DM $\gamma$-ray and shear 
window functions at low redshift, leads to the expectation of a sizable DM 
signal in the angular PS, which is indeed what we find in 
Fig.~\ref{fig:crossexp}.
For $\ell \lesssim 100$ the 2-halo term dominates for all the sources, thus 
the relative size is roughly given by the relative contribution in the 
total EGB emission. At $\ell \gtrsim 100$, the 1-halo term starts to be 
important in the DM case which grows more rapidly than the astrophysical 
sources. At $\ell \gtrsim 10^3$, the 1-halo term takes over also in the SFG 
spectrum which is brought again close to the DM curve. Blazars are largely 
subdominant in the whole range of multipoles.
   
The observational forecasts for the cross-correlation between DES or Euclid 
and Fermi-LAT are shown for the benchmark models 
considered in this work (for error estimates, we take 
observational performances from 
\citep{Atwood:2009ez,Abbott:2005bi,Euclid}). Fig.~\ref{fig:crossexp} shows 
that a DM signal can be disentangled in the angular PS at 
$\ell\lesssim10^3$.
The same conclusion can be derived for DM models with different mass and 
annihilation/decay channels, provided the DM is a significant component 
of the total $\gamma$-ray EGB (at least in one energy bin) 
as in our assumptions.%

%********************************************************************************

\section{Conclusions}
%{\em Conclusions.}
%
In this Letter, we discussed the cross-correlation angular power spectrum of 
weak lensing cosmic shear and $\gamma$-rays produced by WIMP 
annihilations/decays and astrophysical sources.
We showed that this method can provide novel information on the composition of the 
EGB. Since the shear signal is stronger for structures of larger masses and most of the 
$\gamma$-ray emission from decaying and annihilating DM 
is also produced in large mass halos, their cross-correlation is typically 
stronger than the case of astrophysical sources (which are associated to 
galactic-mass halos). The combination of Fermi-LAT with forthcoming surveys like DES and Euclid can thus 
potentially {provide evidence for WIMPs}.

\acknowledgements
%{\em Acknowledgements.} 
%
We thank L. Amendola, A. Cuoco and A. Green for useful comments. 
NF and MR acknowledge INFN grant FA51. NF acknowledges support of Spanish grant MULTIDARK 
CSD2009-00064. MF is supported by a Leverhulme Trust grant. SC 
acknowledges support from FCT-Portugal under grant PTDC/FIS/100170/ 2008. 
SC work is funded by FCT-Portugal under Post-Doctoral Grant 
SFRH/ BPD/ 80274/ 2011.

\end{document}